# Design and Assembly of a Large-aperture Nb$_3$Sn Cos-theta Dipole Coil with Stress Management in Dipole Mirror Configuration

I. Novitski, A.V. Zlobin, E. Barzi, *Senior Member, IEEE,* and D. Turrioni,

*Abstract*—The stress-management cos-theta (SMCT) coil is a new concept which has been proposed and is being developed at Fermilab in the framework of US Magnet Development Program (US-MDP) for high-field and/or large-aperture accelerator magnets based on low-temperature and high-temperature superconductors. The SMCT structure is used to reduce large coil deformations under the Lorentz forces and, thus, the excessively large strains and stresses in the coil. A large-aperture Nb$_3$Sn SMCT dipole coil has been developed and fabricated at Fermilab to demonstrate and test the SMCT concept including coil design, fabrication technology and performance. The first SMCT coil has been assembled with 60-mm aperture Nb$_3$Sn coil inside a dipole mirror configuration and will be tested separately and in series with the insert coil. This paper summarizes the large-aperture SMCT coil design and parameters and reports the coil fabrication steps and its assembly in dipole mirror configuration.

## I. Introduction

**H**IGH-field and/or large-aperture magnets based on Nb$_3$Sn superconductor are needed for future hadron and muon colliders [1], [2]. High magnetic fields cause large mechanical strains and stresses which can degrade or even permanently damage brittle Nb$_3$Sn coils. These forces significantly increase with growing the level of magnetic field and the size of magnet aperture and, at some point, they need special treatment. To stabilize turn position and limit the strain/stress in brittle coil within an acceptable range for the superconductor during magnet fabrication and operation, various stress management coil structures are being developed and studied [3]-[5].

An innovative stress-management (SM) concept for cos-theta (CT) coils (SMCT coil concept) has been proposed and is being developed at Fermilab [5], [6]. A large-aperture Nb$_3$Sn SMCT dipole coil was designed and manufactured to validate and test the SMCT concept including coil design, fabrication technology, and performance. The SMCT coil structure has complex 3D geometry which is hard to make using conventional machining. To overcome these difficulties, advanced Additive Manufacturing (AM) technologies was used for manufacturing of accurate and complex metallic SMCT coil parts [7]. The first SMCT coil was assembled with 60-mm aperture Nb$_3$Sn coil inside a dipole mirror magnet and will be tested separately and in series with the insert coil.

This paper summarizes the SMCT coil design and parameters, describes the coil main fabrication steps, instrumentation, and its assembly in dipole mirror structure. The expected coil performance in different test configurations is presented.

## II. SMCT Coil Design and Technology

### A. SMCT Coil Design

The cross-section of the large-aperture superconducting dipole coil developed at Fermilab based on the SMCT coil geometry is presented in Fig. 1. The SMCT coil consists of 2-layers, each layer has 5 blocks wound into a stainless-steel structure [8]. To produce dipole field in magnet aperture, the number of turns in blocks approximately follows the cos-theta distribution. The coil blocks are separated by 5 mm azimuthally and the coil layers are separated by 5 mm in the radial direction from each other and from the inner coil to provide space for the SMCT support structure. Each coil block in both layers is placed in its specific groove. Consequently, azimuthal and radial components of the Lorentz force in each block are not accumulated but transmitted to the coil and magnet structures bypassing the coil blocks. The coil inner diameter is 123 mm leaving ~0.5 mm of radial space for the coil radial insulation. The coil outer diameter is 206 mm.

The coil uses the 15.1 mm wide and 1.319 mm average thick 40-strand Rutherford keystoned cable with a keystone angle of 0.805 degree. The cable is made of the Nb$_3$Sn composite wire with a Cu/nonCu ratio of 1.13 and $J_c$ at 15 T and 4.2 K of 1500 A/mm$^2$. This cable was developed and fabricated at Fermilab [9] and used in MBHSP [10] and MDPCT1 [11] dipole coils.

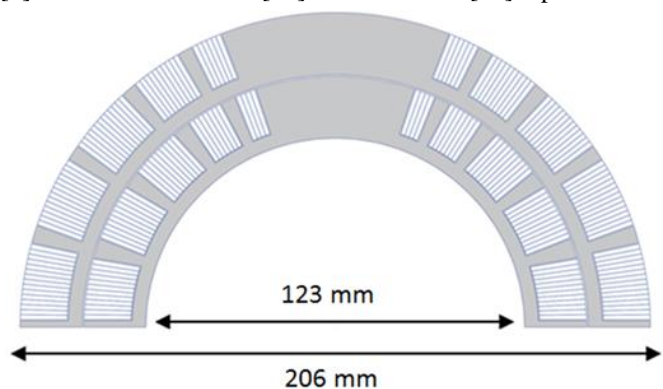

Fig. 1. Cross-section of the large-aperture two-layer SMCT coil.

This work is supported by Fermi Research Alliance, LLC, under contract No. DE-AC02-07CH11359 with the U.S. Department of Energy

Authors are with the Fermi National Accelerator Laboratory (FNAL), Batavia, IL 80510 USA (e-mail: zlobin@fnal.gov).

## B. SMCT Coil Fabrication and Instrumentation

Because of high complexity of the end geometry, the SMCT coil support structure was modeled using plastic parts printed with AM technology and tested in practice winding [12]. The final SMCT coil metallic parts were printed at GE Additive using 316 stainless steel powder. Geometrical parameters of the SMCT coil parts were measured [7]; additional elements, such as special holes, pole slots, etc. were added before coil winding.

Each layer was wound into SMCT coil structure using non-reacted $Nb_3Sn$ cable (Wind-&-React approach). The cable was insulated with 0.075 mm thick E-glass tape in the condition provided by the tape manufacturer, wrapped with ~45% overlap. The SMCT grooves were designed slightly larger to provide room for the block insulation and for the $Nb_3Sn$ cable transverse expansion after reaction. The gaps between coil end and central blocks permitted for cable axial extension throughout coil heating and reaction. The coil was reacted in Argon in the 3-step cycle with $T_{max}=658^oC$ for 48 hours. The SMCT coil pictures after winding and after reaction are shown in Figs. 2 and 3.

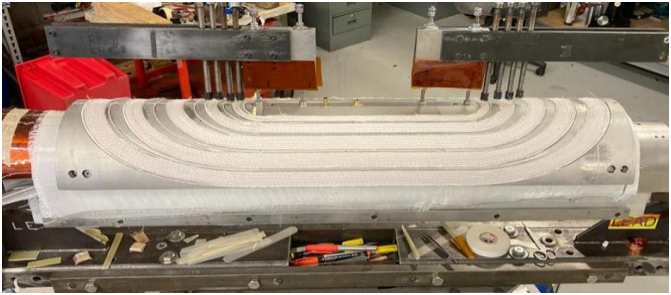

Fig. 2. SMCT coil after winding into SM structure. The space in blocks for cable transverse expansion simplifies insertion of last turns in blocks.

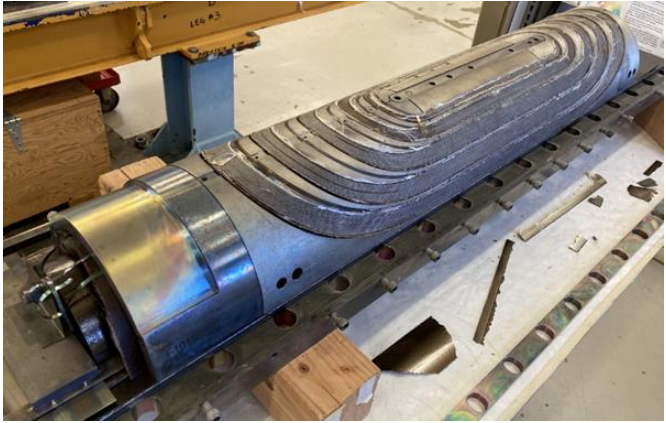

Fig. 3. SMCT coil after reaction. As expected, the space for the cable expansion reduced (but not eliminated). No insulation damage during reaction was found.

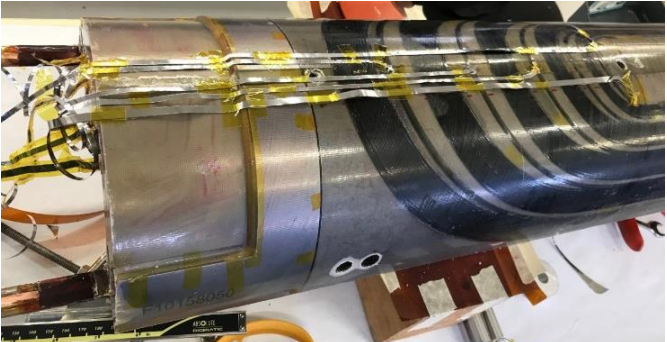

Fig. 4. SMCT coil after impregnation with epoxy resin and VT installation.

After reaction, the flexible Nb-Ti cables were spliced to $Nb_3Sn$ coil leads and covered by the lead end (LE) saddle extension blocks. The coil was wrapped with a 0.125 mm thick S2-glass cloth, potted with CTD101K epoxy resin, and cured at 125°C for 16 hours. A picture of the potted coil with voltage taps (VTs) on the LE outer surface is shown in Fig. 4.

To check the coil geometry after reaction and potting and select the coil shimming plan, the radial and azimuthal sizes of the SMCT coil were measured in several cross-sections along the length with a coordinate measuring machine. Radial variations of ±0.13 mm of the outer surface and ±0.05 mm of the inner surface were measured in the coil straight section. The coil outer surface will be equalized by adding special azimuthal shims. The potting tooling will be improved for the next coils.

To control the coil reaction process and estimate the magnet conductor limits, four witness samples (strands extracted from the cable) were placed inside the reaction fixture - two in the LE and two in the non-LE areas. The $I_c(B)$ plots at T=1.9 K for two tested witness samples are presented in Fig. 5. At 6 T and higher fields the $I_c$ values for both samples are nearly identical. Instabilities in $I_c$ data are seen at fields below 6 T. The level of instability currents is somewhat different for the LE and non-LE samples, but it is well above the coil operation currents.

Measured values of SMCT coil $I_c$ at 15 T are 254 A for the non-LE and 268 A for the LE. These data are in a good agreement with the witness sample tests for the three MDPCT1 dipole outer coils (256, 258 and 266 A) which were made of the same cable ~4 years ago. The *RRR* values of witness samples, are 156 and 101 for the LE and 77 and 74 for the non-LE. These *RRR* data are also consistent with measured *RRR* values which were within 70-100 in two MDPCT1 outer coils [13].

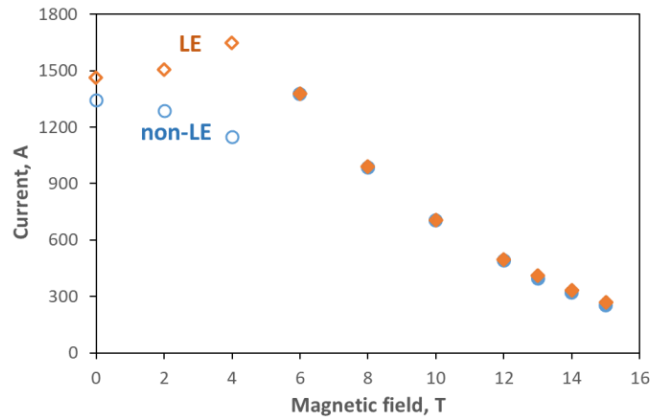

Fig. 5. SMCT coil witness sample $I_c$ vs. *B* at 1.9 K. At 6 T and higher fields the measurement data practically overlap for both samples. Quench currents at low fields induced by the flux-jump instabilities are shown with open symbols.

## C. Mirror Magnet Assembly and Coil Preload

The first SMCT coil will be tested in a dipole mirror configuration using MDPCT1 mechanical structure and two-layer inner coil [11]. The cross-section of four-layer coil assembly with interlayer and ground insulations, quench protection heaters and the stainless-steel outer shell is shown in Fig. 6.

The inner and SMCT coils are divided by three layers of Kapton of a whole thickness of 0.3 mm. Two quench protection heaters, made of 0.025 mm thick stainless-steel strips, were glued to the 0.050 mm thick Kapton tape to cover four largest outer blocks of the SMCT coil. The tape with heaters was placed between the first and the second layer of the four-layer ground insulation. Each layer thickness is 0.125 mm. The total Kapton thickness between the heater and the coil is 0.175 mm. There are no quench protection heaters on the inner coil.

The coil assembly, covered by 1-mm thick 316L stainless steel shell, was placed inside the bottom part of the iron half-yoke. The horizontal yoke split was selected based on the convenience of magnet assembly as well as on the mechanical considerations discussed below. The yoke is made of AISI 1020 iron laminations with the OD of 587 mm, connected by strong 7075-T6 aluminum I-clamps, and enclosed by a 12.5-mm thick 316 stainless-steel skin. To accommodate the SMCT coil with larger OD, the inner diameter of iron laminations was enlarged from 196.1 mm to 209.5 mm. Pictures of the SMCT coil assembly with the iron yoke are shown in Figs. 7-8.

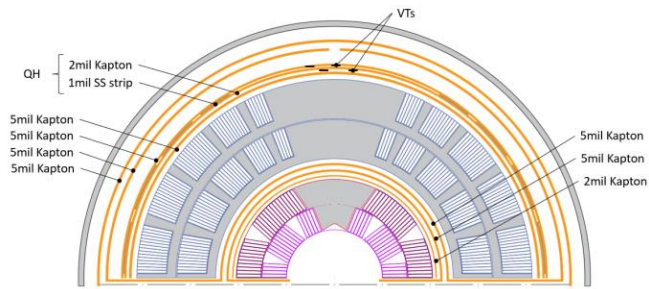

Fig. 6. The cross-section of four-layer coil assembly with interlayer and ground insulation, quench protection heaters and stainless-steel outer shell.

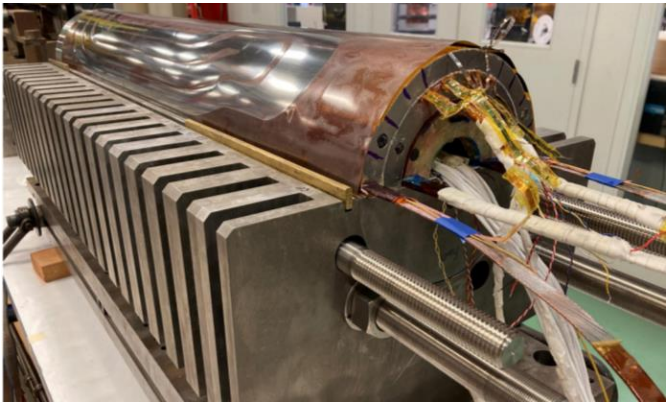

Fig. 7. SMCT coil with inner coil inside its aperture is placed on the iron mirror blocks inside the bottom part of horizontally split iron yoke.

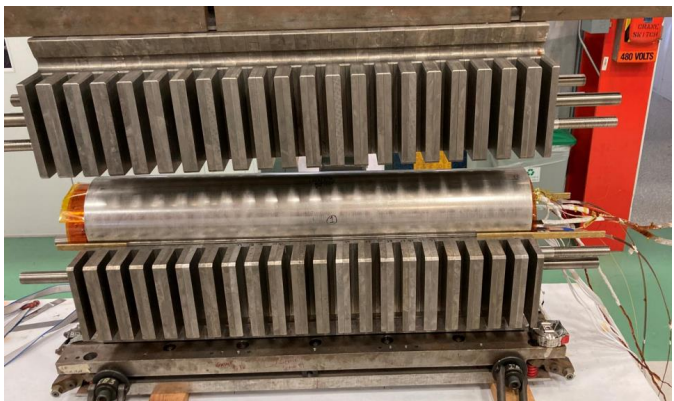

Fig. 8. Coil assembly covered by the protection shell is being enclosed by the top part of the yoke.

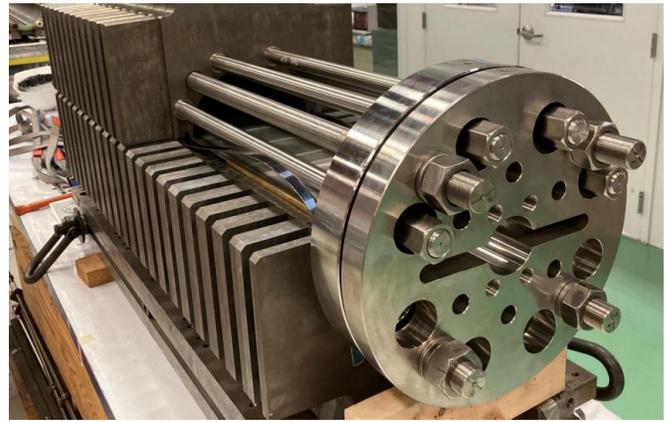

Fig. 9. SMCT coil end support system.

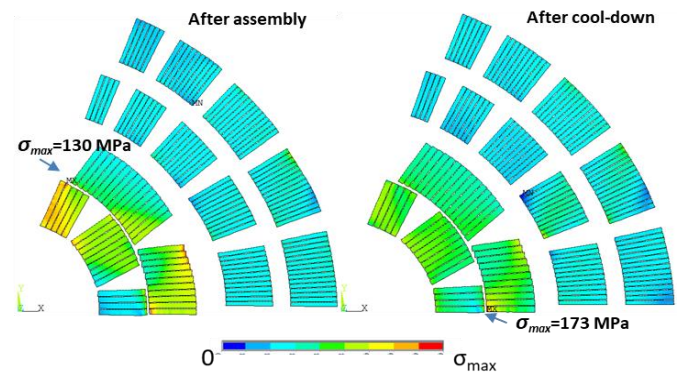

Fig. 10. The calculated variation of the coil stress after assembly (left) and cooling-down to the helium temperature (right). The stress scale at the bottom allows estimating the stress level in various part of the coil.

Coil ends are supported by two independent systems shown in Fig. 9. The outer SMCT coil is supported by eight 24 mm diameter rods and 50 mm thick end plates. The inner coil is supported by four 30 mm diameter rods and 50 mm thick inside and 30 mm thick outside end plates.

The coil pre-load during assembly is created by the coil midplane, intercoil, coil-yoke and the yoke-clamp interferences, and the skin tension after welding. During cooling-down to and at the operation temperature, the coil stress is managed by the gap between the yoke blocks. The calculated coil stress in the dipole mirror with horizontal yoke gap after assembly and at the operation temperature is shown in Fig. 10. The material properties and the details of mechanical analysis for the vertical yoke gap are reported in [12]. With the horizontal yoke gap the maximum stress in the inner coil after assembly is less than 130 MPa and in the SMCT coil is less than 50 MPa. After cooling-down the maximum stress in the inner coil increases to 173 MPa and to 80 MPa in the SMCT coil. Analysis shows that the horizontal yoke gap provides lower stress in the inner coil than the vertical gap [12] making it more attractive.

## III. SMCT Coil Test in Mirror Configuration

### A. Test Configurations and Conductor Limit

The cross-section of the four-layer dipole mirror with outer SMCT coil and the 60-mm aperture inner coil from the 15 T dipole MDPCT1 is shown in Fig. 11 (left).

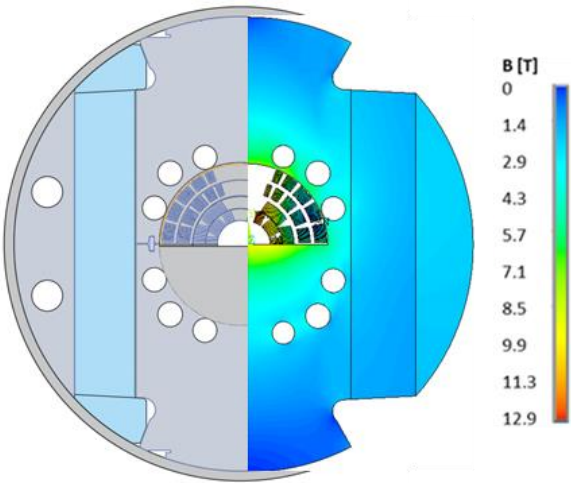

Fig. 11. SMCT coil in the four-layer dipole mirror configuration with the MDPCT1 dipole inner coil and support structure (left) and calculated distribution of magnetic field induction in the coil and iron yoke (right).

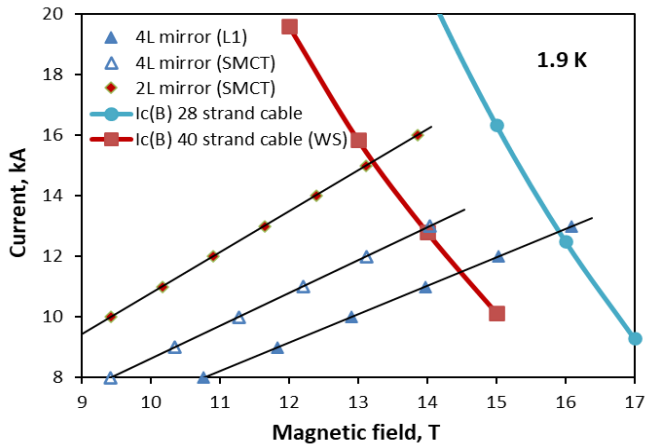

Fig. 12. $I_c(B)$ curves of 40-strand (SMCT coil WS) and 28-strand (MDPCT1 inner coil) Nb$_3$Sn cables at 1.9 K, and load lines of SMCT and insert coils in two-layer and four-layer dipole mirror configurations.

The mirror test will be done in two configurations. First, the SMCT coil will be powered independently, and then in series with the inner coil. In the first configuration, the SMCT coil structure will be tested with respect to the azimuthal Lorentz forces. In the second configuration, the radial Lorentz force component from the inner coil will be added. The distribution of magnetic field in the coil and in the iron yoke cross-sections calculated with ANSYS in the four-layer mirror at the coil current of 10 kA is shown in Fig. 11 (right). Based on calculations, the peak field in the four-layer configuration is in the inner-layer pole turn of the inner coil. The peak field in the SMCT coil is lower and located in the coil inner-layer pole turn.

The $I_c(B)$ curves of the 40-strand and 28-strand Nb$_3$Sn cables based on witness sample data for the SMCT coil and MDPCT1 inner coil, and the corresponding coil load lines in the dipole mirror test configurations are shown in Fig. 12. The calculated short sample limit (SSL) at 1.9 K of the individually powered SMCT coil is 13.2 T at 15.1 kA current. The SSL of the four-layer mirror is defined by the inner coil and is 15.9 T at 12.8 kA current. The peak field in the SMCT coil in the four-layer dipole mirror at the conductor limit is ~13.9 T which is only slightly lower than in the two-layer mirror configuration.

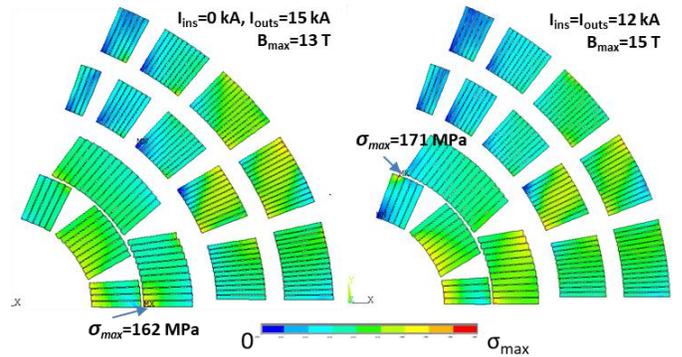

Fig. 13. The calculated variation of the coil stress during powering only the SMCT coil (left) and both coils connected in series (right). The stress scale at the bottom allows estimating the stress level in various part of the coil.

The calculated variations of the equivalent stress in coils at 1.9 K with coil current of 15 kA in two-layer and of 12 kA in four-layer mirrors with horizontally split yoke are reported in Fig. 13. The highest stresses in the coil for both two-layer and four-layer mirror arrangements are achieved in the inner coil on the level of 160-170 MPa. The stress distribution and its level also look better for the horizontally split yoke than for the yoke with vertical gap [12]. Mechanical calculations confirms that the highest stresses in the coils and in the dipole mirror structure are within the tolerable ranges. During powering the coil turns are shifted under Lorentz forces towards the coil mid-planes. The impact of these gaps on magnet training and degradation will be investigated experimentally during mirror magnet test.

## IV. Conclusion

The SMCT coil concept was proposed at Fermilab for high-field and/or large-aperture accelerator magnets based on low-temperature and high-temperature superconductors [14], [15]. The SMCT coil structure permits reducing coil deformations under the Lorentz forces and, thus, the excessively high stresses in the coil and a separation of pole turns at high fields.

A large-aperture Nb$_3$Sn SMCT dipole coil was designed and built at Fermilab to validate and study the SMCT coil concept. The SMCT coil design and manufacturing technology were tested and optimized using 3D printed plastic parts and practice coil winding. The first set of SMCT coil parts for Nb$_3$Sn dipole coil was printed by GE Additive using 316 stainless-steel powder. The first 123-mm aperture SMCT coil was manufactured, measured, instrumented, and assembled in a dipole mirror configuration with a two-layer 60-mm aperture inner coil used in MDPCT1 dipole. The SMCT coil will be tested independently, and then in series with the inner coil. The coil electrical and mechanical performance parameters in two configurations were analyzed and will be compared with experimental data. Mirror magnet tests are planned for March-April 2023.

## Acknowledgments

The authors thank Vadim Kashikhin for the magnetic optimization of the coil cross-section, Jodi Coghill for the end part design, Allen Rusy and James Karambis for the technical support of this work.

## REFERENCES

[1] L. Bottura, S. Gourlay, A. Yamamoto, A.V. Zlobin, "Superconducting Magnets for Particle Accelerators", *IEEE Trans. on Nuclear Science*, vol. 63, no. 2, Apr. 2016.

[2] L. Bottura, S. Prestemon, L. Rossi, A.V. Zlobin, "Magnet Designs and Technologies for Future Colliders," Frontiers in Physics, 2022, 0:935196.

[3] T. Elliott *et al.*, "16 Tesla $Nb_3Sn$ Dipole Development at Texas A&M University," *IEEE Trans. on Appl. Supercond.*, vol. 7, no. 2, June 1997, p. 555.

[4] S. Caspi *et al.*, "Canted-Cosine-Theta Magnet (CCT) – A Concept for High Field Accelerator Magnets," *IEEE Trans. on Appl. Supercond.*, vol. 24, no. 3, June 2014, Art. no. 4001804.

[5] V.V. Kashikhin, I. Novitski, A.V. Zlobin, "Design studies and optimization of a high-field dipole for a future Very High Energy *pp* Collider", *Proc. of IPAC2017*, Copenhagen, Denmark, May 2017, p. 3597.

[6] A.V. Zlobin et al., "Conceptual design of a 17 T $Nb_3Sn$ accelerator dipole magnet," *Proc. of IPAC2018*, WEPML027, 2018, p. 2742.

[7] I Novitski et al., "Using Additive Manufacturing technologies in high-field accelerator magnet coils," *CEC-ICMC'21*, FERMILAB-CONF-21-369-TD, 2021.

[8] I. Novitski et al., "Development of a 15 T $Nb_3Sn$ Accelerator Dipole Demonstrator at Fermilab," *IEEE Trans. on Appl. Supercond.*, vol. 26, no. 4, Jun. 2016, Art. no. 4001007.

[9] E. Barzi et al., "Development and Fabrication of $Nb_3Sn$ Rutherford Cable for the 11 T DS Dipole Demonstration Model", *IEEE Trans. on Appl. Supercond.*, vol. 22, no. 3, June 2012, Art. no. 6000805

[10] A. V. Zlobin et al., "Development and test of a single-aperture 11T $Nb_3Sn$ demonstrator dipole for LHC upgrades", *IEEE Trans. on Appl. Supercond.*, vol. 23, no. 3, June 2013, Art. no. 4000904.

[11] A.V. Zlobin et al., "Development and First Test of the 15 T $Nb_3Sn$ Dipole Demonstrator MDPCT1", *IEEE Trans. on Appl. Supercond.*, vol. 30, no. 4, 2020, 10.1109/TASC.2020.2967686

[12] I. Novitski et al., "Development of 120-mm diameter $Nb_3Sn$ dipole coil with stress management", *IEEE Trans. on Appl. Supercond.*, vol. 32, no. 6, Sept. 2022, Art. no. 4006005.

[13] A.V. Zlobin et al., "Reassembly and Test of High-Field $Nb_3Sn$ Dipole Demonstrator MDPCT1", *IEEE Trans. on Appl. Supercond.*, vol. 31, no. 5, August 2021, 10.1109/TASC.2020.2967686

[14] A.V. Zlobin, V.V. Kashikhin, I. Novitski, "Large-aperture high-field $Nb_3Sn$ dipole magnets," *Proc. of IPAC2018*, WEPML026, 2018, p. 2738.

[15] A.V. Zlobin, I. Novitski and E. Barzi, "Conceptual Design of a HTS Dipole Insert Based on Bi2212 Rutherford Cable," *Instruments* 2020, 4, 29.